# Fractal Fluctuations and Statistical Normal Distribution

## A. M. Selvam

Deputy Director (Retired)
Indian Institute of Tropical Meteorology, Pune 411 008, India
Email: amselvam@gmail.com
Web sites: http://www.geocities.com/amselvam
http://amselvam.tripod.com/index.html

Dynamical systems in nature exhibit selfsimilar fractal fluctuations and the corresponding power spectra follow inverse power law form signifying long-range space-time correlations identified as self-organized criticality. The physics of self-organized criticality is not yet identified. The Gaussian probability distribution used widely for analysis and description of large data sets underestimates the probabilities of occurrence of extreme events such as stock market crashes, earthquakes, heavy rainfall, etc. The assumptions underlying the normal distribution such as fixed mean and standard deviation, independence of data, are not valid for real world fractal data sets exhibiting a scale-free power law distribution with fat tails. A general systems theory for fractals visualizes the emergence of successively larger scale fluctuations to result from the space-time integration of enclosed smaller scale fluctuations. The model predicts a universal inverse power law incorporating the *golden mean* for fractal fluctuations and for the corresponding power spectra, i.e., the variance spectrum represents the probabilities, a signature of quantum systems. Fractal fluctuations therefore exhibit quantum-like chaos. The model predicted inverse power law is very close to the Gaussian distribution for small-scale fluctuations, but exhibits a *fat long tail* for large-scale fluctuations. Extensive data sets of Dow Jones index, Human DNA, Takifugu rubripes (Puffer fish) DNA are analysed to show that the space/time data sets are close to the model predicted power law distribution.

## 1. Introduction

Dynamical systems in nature such as atmospheric flows, heartbeat patterns, population dynamics, stock market indices, DNA base A, C, G, T sequence pattern, etc., exhibit irregular space-time fluctuations on all scales and exact quantification of the fluctuation pattern for predictability purposes has not yet been achieved. Traditionally, the Gaussian probability distribution is used for a broad quantification of the data set variability in terms of the sample mean and variance. The fractal or selfsimilar nature of space-time fluctuations was identified by Mandelbrot (1975) in the 1970s. Representative examples of fractal fluctuations of (i) Daily percentage change of Dow Jones Index (ii) Human DNA base CG concentration/10bp (base pairs) (iii) Takifugu rubripes (Puffer fish) DNA base CG concentration/10bp are shown in Fig. 1.

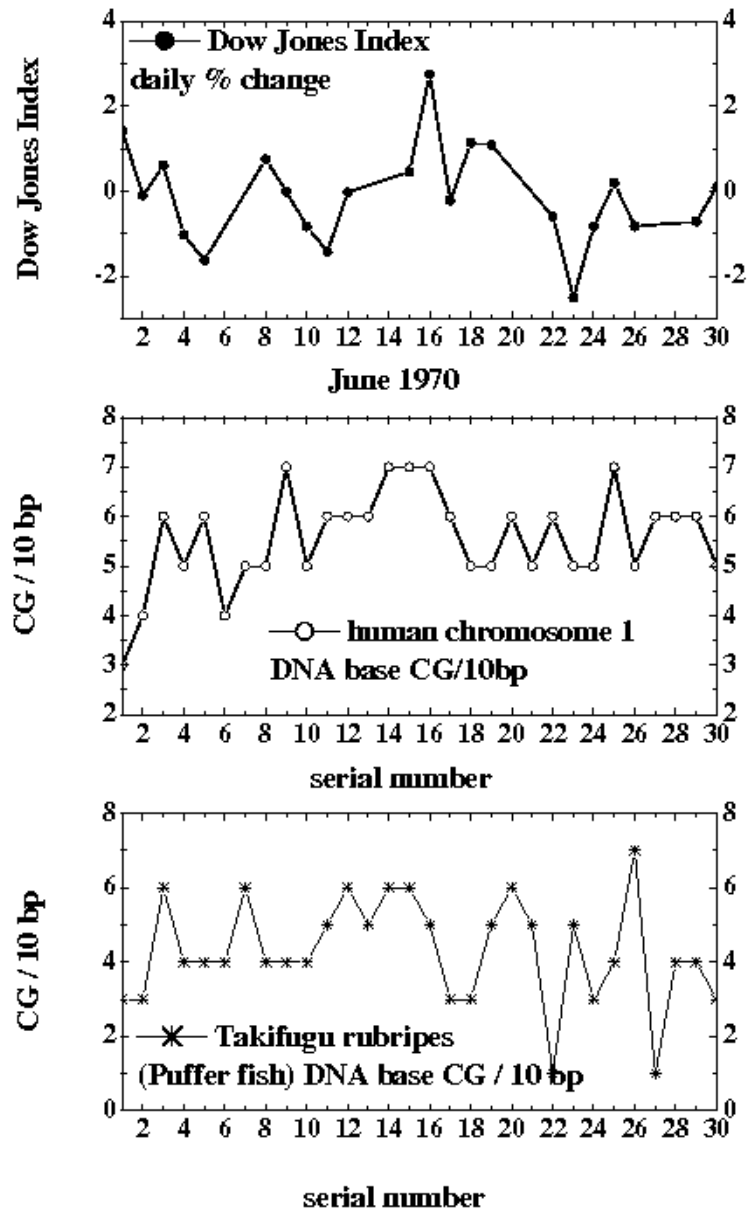

Figure 1: Representative examples of fractal fluctuations of (i) Daily percentage change of Dow Jones Index (ii) Human DNA base CG concentration/10bp (base pairs) (iii) Takifugu rubripes (Puffer fish) DNA base CG concentration/10bp

Fractal fluctuations (Fig. 1) show a zigzag selfsimilar pattern of successive increase followed by decrease on all scales (space-time), for example in atmospheric flows, cycles of increase and decrease in meteorological parameters such as wind, temperature, etc. occur from the turbulence scale of millimeters-seconds to climate scales of thousands of kilometers-years. The power spectra of fractal fluctuations exhibit inverse power law of the form $f^{\alpha}$ where $f$ is the frequency and $\alpha$ is a constant.

Inverse power law for power spectra indicate long-range space-time correlations or scale invariance for the scale range for which α is a constant, i.e., the amplitudes of the eddy fluctuations in this scale range are a function of the scale factor α alone. In general the value of α is different for different scale ranges indicating multifractal structure for the fluctuations. The long-range space-time correlations exhibited by dynamical systems are identified as self-organized criticality (Bak *et al*., 1988; Schroeder, 1990). The physics of self-organized criticality is not yet identified. The physics of fractal fluctuations generic to dynamical systems in nature is not yet identified and traditional statistical, mathematical theories do not provide adequate tools for identification and quantitative description of the observed universal properties of fractal structures observed in all fields of science and other areas of human interest. A recently developed general systems theory for fractal space-time fluctuations (Selvam, 1990, 2005, 2007; Selvam and Fadnavis, 1998) shows that the larger scale fluctuation can be visualized to emerge from the space-time averaging of enclosed small scale fluctuations, thereby generating a hierarchy of selfsimilar fluctuations manifested as the observed eddy continuum in power spectral analyses of fractal fluctuations. Such a concept results in inverse power law form incorporating the golden mean τ for the space-time fluctuation pattern and also for the power spectra of the fluctuations (Sec. 3). The predicted distribution is close to the Gaussian distribution for small-scale fluctuations, but exhibits *fat long tail* for large-scale fluctuations. Analysis of extensive data sets of (i) Daily percentage change of Dow Jones Index (ii) Human DNA base CG concentration/10bp (base pairs) (iii) Takifugu rubripes (Puffer fish) DNA base CG concentration/10bp show that the data follow closely, but not exactly the statistical normal distribution, particularly in the region of normalized deviations *t* greater than 2, the *t* values being computed as equal to ($x$-$av$)/$sd$ where $av$ and $sd$ denote respectively the mean and standard deviation of the variable *x*. The general systems theory, originally developed for turbulent fluid flows, provides universal quantification of physics underlying fractal fluctuations and is applicable to all dynamical systems in nature independent of its physical, chemical, electrical, or any other intrinsic characteristic. In the following, Sec. 2 gives a summary of traditional statistical and mathematical theories/techniques used for analysis and quantification of space-time fluctuation data sets. The general systems theory for fractal space-time fluctuations is described in Sec. 3. Sec. 4 deals with data and analyses techniques. Discussion and conclusions of results are presented in Sec. 5.

## 2. Statistical methods for data analysis

Dynamical systems such as atmospheric flows, stock markets, heartbeat patterns, population growth, traffic flows, etc., exhibit irregular space-time fluctuation patterns. Quantification of the space-time fluctuation pattern will help predictability studies, in particular for events which affect day-to-day human life such as extreme weather events, stock market crashes, traffic jams, etc. The analysis of data sets and broad quantification in terms of probabilities belongs to the field of statistics. Early attempts resulted in identification of the following two quantitative (mathematical) distributions which approximately fit data sets from a wide range of scientific and other disciplines of study. The first is the well known statistical normal distribution and the second is the power law distribution associated with the recently identified 'fractals' or selfsimilar characteristic of data sets in general. In the following, a summary is given of the history and merits of the two distributions.

## 2.1 Statistical normal distribution

Historically, our present day methods of handling experimental data have their roots about four hundred years ago. At that time scientists began to calculate the odds in gambling games. From those studies emerged the theory of probability and subsequently the theory of statistics. These new statistical ideas suggested a different and more powerful experimental approach. The basic idea was that in some experiments random errors would make the value measured a bit higher and in other experiments random errors would make the value measured a bit lower. Combining these values by computing the average of the different experimental results would make the errors cancel and the average would be closer to the "right" value than the result of any one experiment (Liebovitch and Scheurle, 2000).

Abraham de Moivre, an 18th century statistician and consultant to gamblers made the first recorded discovery of the normal curve of error (or the bell curve because of its shape) in 1733. The normal distribution is the limiting case of the binomial distribution resulting from random operations such as flipping coins or rolling dice. Serious interest in the distribution of errors on the part of mathematicians such as Laplace and Gauss awaited the early nineteenth century when astronomers found the bell curve to be a useful tool to take into consideration the errors they made in their observations of the orbits of the planets (Goertzel and Fashing, 1981, 1986). The importance of the normal curve stems primarily from the fact that the distributions of many natural phenomena are at least approximately normally distributed. This normal distribution concept has molded how we analyze experimental data over the last two hundred years. We have come to think of data as having values most of which are near an average value, with a few values that are smaller, and a few that are larger. The probability density function, PDF($x$), is the probability that any measurement has a value between $x$ and $x + dx$. We suppose that the PDF of the data has a normal distribution. Most quantitative research involves the use of statistical methods presuming *independence* among data points and Gaussian 'normal' distributions (Andriani and McKelvey, 2007). The Gaussian distribution is reliably characterized by its stable mean and finite variance (Greene, 2002). Normal distributions place a trivial amount of probability far from the mean and hence the mean is representative of most observations. Even the largest deviations, which are exceptionally rare, are still only about a factor of two from the mean in either direction and are well characterized by quoting a simple standard deviation (Clauset, Shalizi, and Newman, 2007). However, apparently rare real life catastrophic events such as major earth quakes, stock market crashes, heavy rainfall events, etc., occur more frequently than indicated by the normal curve, i.e., they exhibit a probability distribution with a *fat tail*. Fat tails indicate a power law pattern and interdependence. The "tails" of a power-law curve — the regions to either side that correspond to large fluctuations — fall off very slowly in comparison with those of the bell curve (Buchanan, 2004). The normal distribution is therefore an inadequate model for extreme departures from the mean.

The following references are cited by Goertzel and Fashing (1981, 1986) to show that the bell curve is an empirical model without supporting theoretical basis: (i) Modern texts usually recognize that there is no theoretical justification for the use of the normal curve, but justify using it as a convenience (Cronbach, 1970). (ii) The bell curve came to be generally accepted, as M. Lippmnan remarked to Poincare (Bradley, 1969), because "...the experimenters fancy that it is a theorem in mathematics and the mathematicians that it is an experimental fact". (iii) Karl Pearson (best known today for the invention of the product-moment correlation coefficient) used his newly developed Chi Square test to check how closely a number of empirical distributions of supposedly random errors fitted the bell curve. He found that many of the distributions that had been cited in the literature as fitting the normal curve were actually significantly different from it, and concluded that "the normal curve of error possesses no special fitness for describing errors or deviations such as arise either in observing practice or in nature" (Pearson, 1900).

## 2.2 Fractal fluctuations and statistical analysis

Fractals are the latest development in statistics. The space-time fluctuation pattern in dynamical systems was shown to have a selfsimilar or fractal structure in the 1970s (Mandelbrot, 1975). The larger scale fluctuation consists of smaller scale fluctuations identical in shape to the larger scale. An appreciation of the properties of fractals is changing the most basic ways we analyze and interpret data from experiments and is leading to new insights into understanding physical, chemical, biological, psychological, and social systems. Fractal systems extend over many scales and so cannot be characterized by a single characteristic average number (Liebovitch and Scheurle, 2000). Further, the selfsimilar fluctuations imply long-range space-time correlations or interdependence. Therefore, the Gaussian distribution will not be applicable for description of fractal data sets. However, the bell curve still continues to be used for approximate quantitative characterization of data which are now identified as fractal space-time fluctuations.

### *2.2.1 Power laws and fat tails*

Fractals conform to power laws. A power law is a relationship in which one quantity *A* is proportional to another *B* taken to some power *n*; that is, $A \sim B^n$ (Buchanan, 2004). One of the oldest scaling laws in geophysics is the Omori law (Omori, 1895). This law describes the temporal distribution of the number of after-shocks, which occur after a larger earthquake (i.e. main-shock) by a scaling relationship. Richardson (1960) came close to the concept of fractals when he noted that the estimated length of an irregular coastline scales with the length of the measuring unit. Andriani and McKelvey (2007) have given exhaustive references to earliest known work on power law relationships summarized as follows. Pareto (1897) first noticed power laws and fat tails in economics. Cities follow a power law when ranked by population (Auerbach, 1913). Dynamics of earthquakes follow power law (Gutenberg and Richter, 1944) and Zipf (1949) found that a power law applies to word frequencies (Estoup (1916), had earlier found a similar relationship). Mandelbrot (1963) rediscovered them in the 20th century, spurring a small wave of interest in finance (Fama, 1965; Montroll and Shlesinger, 1984). However, the rise of the 'standard' model (Gaussian) of efficient markets, sent power law models into obscurity. This lasted until the 1990s, when the occurrence of catastrophic events, such as the 1987 and 1998 financial crashes, that were difficult to explain with the 'standard' models (Bouchaud *et al*., 1998), re-kindled the fractal model (Mandelbrot and Hudson, 2004).

A power law world is dominated by extreme events ignored in a Gaussian-world. In fact, the fat tails of power law distributions make large extreme events orders-of-magnitude more likely. Theories explaining power laws are also scale-free. This is to say, the same explanation (theory) applies at all levels of analysis (Andriani and McKelvey, 2007).

*2.2.2 Scale-free theory for power laws with fat, long tails*

A scale-free theory for the observed fractal fluctuations in atmospheric flows shows that the observed long-range spatiotemporal correlations are intrinsic to quantumlike chaos governing fluid flows. The model concepts are independent of the exact details such as the chemical, physical, physiological and other properties of the dynamical system and therefore provide a general systems theory applicable to all real world and computed dynamical systems in nature (Selvam, 1998, 1999, 2001a, b, 2002a, b, 2004, 2005, 2007; Selvam *et al*., 2000). The model is based on the concept that the irregular fractal fluctuations may be visualized to result from the superimposition of an eddy continuum, i.e., a hierarchy of eddy circulations generated at each level by the space-time integration of enclosed small-scale eddy fluctuations. Such a concept of space-time fluctuation averaged distributions *should* follow statistical normal distribution according to *Central Limit Theorem* in traditional Statistical theory (Ruhla, 1992). Also, traditional statistical/mathematical theory predicts that the Gaussian, its Fourier transform and therefore Fourier transform associated power spectrum are the same distributions. The Fourier transform of normal distribution is essentially a normal distribution. A power spectrum is based on the Fourier transform, which expresses the relationship between time (space) domain and frequency domain description of any physical process (Phillips, 2005; Riley, Hobson and Bence, 2006). However, the model (Sec. 3) visualises the eddy growth process in successive stages of unit length-step growth with ordered two-way energy feedback between the larger and smaller scale eddies and derives a power law probability distribution *P* which is close to the Gaussian for small deviations and gives the observed fat, long tail for large fluctuations. Further, the model predicts the power spectrum of the eddy continuum also to follow the power law probability distribution *P*.

In summary, the model predicts the following: (i) The eddy continuum consists of an overall logarithmic spiral trajectory with the quasiperiodic *Penrose* tiling pattern for the internal structure. (ii)The successively larger eddy space-time scales follow the Fibonacci number series. (iii) The probability distribution *P* of fractal domains for the $n^{th}$ step of eddy growth is equal to $\tau^{-4n}$ where $\tau$ is the golden mean equal to $(1+\sqrt{5})/2$ ($\approx 1.618$). The probability distribution *P* is close to the statistical normal distribution for *n* values less than 2 and greater than normal distribution for *n* more than 2, thereby giving a *fat, long tail*. (iv) The probability distribution *P* also represents the relative eddy energy flux in the fractal domain. The square of the eddy amplitude (variance) represents the eddy energy and therefore the eddy probability density *P*. Such a result that the additive amplitudes of eddies when squared represent probabilities, is exhibited by the sub-atomic dynamics of quantum systems such as the electron or proton (Maddox, 1988, 1993; Rae, 1988). Therefore fractal fluctuations are signatures of quantumlike chaos in dynamical systems. (v) The universal algorithm for self-organized criticality is expressed in terms of the universal *Feigenbaum*'s constants (Feigenbaum, 1980) *a* and *d* as $2a^2 = \pi d$ where the fractional volume intermittency of occurrence $\pi d$ contributes to the total variance $2a^2$ of fractal structures. (vi) The *Feigenbaum*'s constants are expressed as functions of the *golden mean*. The probability distribution *P* of fractal domains is also expressed in terms of the Feigenbaum's constants *a* and *d*. The details of the model are summarized in the following section (Sec. 3)

## 3. A general systems theory for fractal fluctuations

The fractal space-time fluctuations of dynamical systems may be visualized to result from the superimposition of an ensemble of eddies (sine waves), namely an eddy continuum. The relationship between large and small eddy circulation parameters are obtained on the basis of Townsend's (1956) concept that large eddies are envelopes enclosing turbulent eddy (small-scale) fluctuations (Fig. 2).

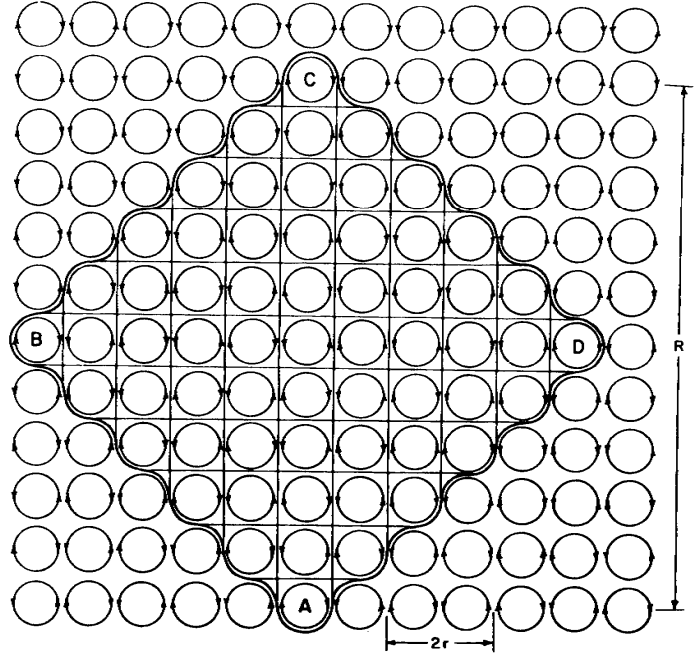

Figure 2: Physical concept of eddy growth process by the self-sustaining process of ordered energy feedback between the larger and smaller scales, the smaller scales forming the internal circulations of larger scales. The figure shows a uniform distribution of dominant turbulent scale eddies of length scale 2r. Larger-eddy circulations such as ABCD form as coherent structures sustained by the enclosed turbulent eddies.

The relationship between root mean square (r. m. s.) circulation speeds $W$ and $w_*$ respectively of large and turbulent eddies of respective radii $R$ and $r$ is then given as

$$W^2 = \frac{2}{\pi}\frac{r}{R}w_*^2 \qquad (1)$$

The dynamical evolution of space-time fractal structures is quantified in terms of ordered energy flow between fluctuations of all scales in Eq. (1), because the square of the eddy circulation speed represents the eddy energy (kinetic). A hierarchical continuum of eddies is generated by the integration of successively larger enclosed turbulent eddy circulations. Such a concept of space-time fluctuation averaged distributions *should* follow statistical normal distribution according to *Central Limit Theorem* in traditional Statistical theory (Ruhla, 1992). Also, traditional statistical/mathematical theory predicts that the Gaussian, its Fourier transform and therefore Fourier transform associated power spectrum are the same distributions.

However, the general systems theory (Selvam, 1998, 1999, 2001a, b, 2002a, b, 2004, 2005, 2007; Selvam *et al.*, 2000) visualises the eddy growth process in successive stages of unit length-step growth with ordered two-way energy feedback between the larger and smaller scale eddies and derives a power law probability distribution *P* which is close to the Gaussian for small deviations and gives the observed fat, long tail for large fluctuations. Further, the model predicts the power spectrum of the eddy continuum also to follow the power law probability distribution *P*. Therefore the additive amplitudes of the eddies when squared (variance), represent the probability distribution similar to the subatomic dynamics of quantum systems such as the electron or photon. Fractal fluctuations therefore exhibit quantumlike chaos.

The above-described analogy of quantumlike mechanics for dynamical systems is similar to the concept of a subquantum level of fluctuations whose space-time organization gives rise to the observed manifestation of subatomic phenomena, i.e., quantum systems as order out of chaos phenomena (Grossing, 1989).

### 3.1 Quasicrystalline structure of the eddy continuum

The turbulent eddy circulation speed and radius increase with the progressive growth of the large eddy (Selvam, 1990). The successively larger turbulent fluctuations, which form the internal structure of the growing large eddy, may be computed (Eq. 1) as

$$w_*^2 = \frac{\pi}{2} \frac{R}{dR} W^2 \qquad (2)$$

During each length step growth $dR$, the small-scale energizing perturbation $W_n$ at the $n^{th}$ instant generates the large-scale perturbation $W_{n+1}$ of radius $R$ where $R = \sum_1^n dR$ since successive length-scale doubling gives rise to $R$. Eq. 2 may be written in terms of the successive turbulent circulation speeds $W_n$ and $W_{n+1}$ as

$$W_{n+1}^2 = \frac{\pi}{2} \frac{R}{dR} W_n^2 \qquad (3)$$

The angular turning $d\theta$ inherent to eddy circulation for each length step growth is equal to $dR/R$. The perturbation $dR$ is generated by the small-scale acceleration $W_n$ at any instant $n$ and therefore $dR=W_n$. Starting with the unit value for $dR$ the successive $W_n$, $W_{n+1}$, $R$, and $d\theta$ values are computed from Eq. 3 and are given in Table 2.

Table 2. The computed spatial growth of the strange-attractor design traced by the macro-scale dynamical system of atmospheric flows as shown in Fig. 3.

| $R$ | $W_n$ | $dR$ | $d\theta$ | $W_{n+1}$ | $\theta$ |
|---|---|---|---|---|---|
| 1.000 | 1.000 | 1.000 | 1.000 | 1.254 | 1.000 |
| 2.000 | 1.254 | 1.254 | 0.627 | 1.985 | 1.627 |
| 3.254 | 1.985 | 1.985 | 0.610 | 3.186 | 2.237 |
| 5.239 | 3.186 | 3.186 | 0.608 | 5.121 | 2.845 |
| 8.425 | 5.121 | 5.121 | 0.608 | 8.234 | 3.453 |
| 13.546 | 8.234 | 8.234 | 0.608 | 13.239 | 4.061 |
| 21.780 | 13.239 | 13.239 | 0.608 | 21.286 | 4.669 |
| 35.019 | 21.286 | 21.286 | 0.608 | 34.225 | 5.277 |
| 56.305 | 34.225 | 34.225 | 0.608 | 55.029 | 5.885 |
| 90.530 | 55.029 | 55.029 | 0.608 | 88.479 | 6.493 |

It is seen that the successive values of the circulation speed $W$ and radius $R$ of the growing turbulent eddy follow the Fibonacci mathematical number series such that $R_{n+1}=R_n+R_{n-1}$ and $R_{n+1}/R_n$ is equal to the golden mean $\tau$, which is equal to $[(1 + \sqrt{5})/2] \cong (1.618)$. Further, the successive $W$ and $R$ values form the geometrical progression $R_0(1+\tau+\tau^2+\tau^3+\tau^4+ ....)$ where $R_0$ is the initial value of the turbulent eddy radius.

Turbulent eddy growth from primary perturbation $OR_O$ starting from the origin O (Fig. 3) gives rise to compensating return circulations $OR_1R_2$ on either side of $OR_O$, thereby generating the large eddy radius $OR_1$ such that $OR_1/OR_O=\tau$ and $R_OOR_1=\pi/5=R_OR_1O$. Therefore, short-range circulation balance requirements generate successively larger circulation patterns with precise geometry that is governed by the *Fibonacci* mathematical number series, which is identified as a signature of the universal period doubling route to chaos in fluid flows, in particular atmospheric flows. It is seen from Fig. 3 that five such successive length step growths give successively increasing radii $OR_1$, $OR_2$, $OR_3$, $OR_4$ and $OR_5$ tracing out one complete vortex-roll circulation such that the scale ratio $OR_5/OR_O$ is equal to $\tau^5=11.1$. The envelope $R_1R_2R_3R_4R_5$ (Fig. 3) of a dominant large eddy (or vortex roll) is found to fit the logarithmic spiral $R=R_0e^{b\theta}$ where $R_0=OR_O$, $b=\tan\delta$ with $\delta$ the crossing angle equal to $\pi/5$, and the angular turning $\theta$ for each length step growth is equal to $\pi/5$. The successively larger eddy radii may be subdivided again in the golden mean ratio. The internal structure of large-eddy circulations is therefore made up of balanced small-scale circulations tracing out the well-known quasi-periodic *Penrose* tiling pattern identified as the quasi-crystalline structure in condensed matter physics. A complete description of the atmospheric flow field is given by the quasi-periodic cycles with *Fibonacci* winding numbers.

### 3.2 Model predictions

The model predictions (Selvam, 1990, 2005, 2007; Selvam and Fadnavis, 1998) are

(a) Atmospheric flows trace an overall logarithmic spiral trajectory $OR_OR_1R_2R_3R_4R_5$ simultaneously in clockwise and anti-clockwise directions with the quasi-periodic *Penrose tiling pattern* (Steinhardt, 1997) for the internal structure shown in Fig. 3.

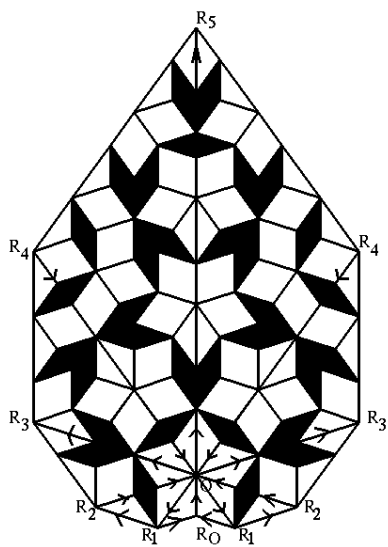

Figure 3. The quasiperiodic *Penrose* tiling pattern

The spiral flow structure can be visualized as an eddy continuum generated by successive length step growths $OR_O$, $OR_1$, $OR_2$, $OR_3$,....respectively equal to $R_1$, $R_2$, $R_3$,....which follow *Fibonacci* mathematical series such that $R_{n+1}=R_n+R_{n-1}$ and $R_{n+1}/R_n=\tau$ where $\tau$ is the *golden mean* equal to $(1+\sqrt{5})/2$ ($\approx 1.618$). Considering a normalized length step equal to 1 for the last stage of eddy growth, the successively decreasing radial length steps can be expressed as 1, $1/\tau$, $1/\tau^2$, $1/\tau^3$, ......The normalized eddy continuum comprises of fluctuation length scales 1, $1/\tau$, $1/\tau^2$, ........ The probability of occurrence is equal to $1/\tau$ and $1/\tau^2$ respectively for eddy length scale $1/\tau$ in any one or both rotational (clockwise and anti-clockwise) directions.

Eddy fluctuation length of amplitude $1/\tau$ has a probability of occurrence equal to $1/\tau^2$ in both rotational directions, i.e., the square of eddy amplitude represents the probability of occurrence in the eddy continuum. Similar result is observed in the subatomic dynamics of quantum systems which are visualized to consist of the superimposition of eddy fluctuations in wave trains (eddy continuum).

(b) The overall logarithmic spiral flow structure is given by the relation

$$W = \frac{w_*}{k} \ln z \qquad (4)$$

In Eq. (4) the constant $k$ is the steady state fractional volume dilution of large eddy by inherent turbulent eddy fluctuations and $z$ is the length scale ratio $R/r$. The constant $k$ is equal to $1/\tau^2$ ($\cong 0.382$) and is identified as the universal constant for deterministic chaos in fluid flows. The steady state emergence of fractal structures is therefore equal to

$$\frac{1}{k} \cong 2.62 \qquad (5)$$

In Eq. (4), $W$ represents the standard deviation of eddy fluctuations, since $W$ is computed as the instantaneous r. m. s. (root mean square) eddy perturbation amplitude with reference to the earlier step of eddy growth. For two successive stages of eddy growth starting from primary perturbation $w_*$, the ratio of the standard deviations $W_{n+1}$ and $W_n$ is given from Eq. (4) as $(n+1)/n$. Denoting by $\sigma$ the standard deviation of eddy fluctuations at the reference level ($n=1$) the standard deviations of eddy fluctuations for successive stages of eddy growth are given as integer multiples of $\sigma$, i.e., $\sigma$, $2\sigma$, $3\sigma$, etc. and correspond respectively to

$$\text{statistical normalised standard deviation } t = 0, 1, 2, 3, .... \qquad (6)$$

The conventional power spectrum plotted as the variance versus the frequency in log-log scale will now represent the eddy probability density on logarithmic scale versus the standard deviation of the eddy fluctuations on linear scale since the logarithm of the eddy wavelength represents the standard deviation, i.e., the r. m. s. value of eddy fluctuations (Eq. 4). The r. m. s. value of eddy fluctuations can be represented in terms of statistical normal distribution as follows. A normalized standard deviation $t=0$ corresponds to cumulative percentage probability density equal to 50 for the mean value of the distribution. Since the logarithm of the wavelength represents the r. m. s. value of eddy fluctuations the normalized standard deviation $t$ is defined for the eddy energy as

$$t = \frac{\log L}{\log T_{50}} - 1 \qquad (7)$$

In Eq. (7) $L$ is the time period (or wavelength) and $T_{50}$ is the period up to which the cumulative percentage contribution to total variance is equal to 50 and $t = 0$. $\log T_{50}$ also represents the mean value for the r. m. s. eddy fluctuations and is consistent with the concept of the mean level represented by r. m. s. eddy fluctuations. Spectra of time series of meteorological parameters when plotted as cumulative percentage contribution to total variance versus *t* have been shown to follow the model predicted universal spectrum (Selvam and Fadnavis, 1998) which is identified as a signature of quantumlike chaos.

(c) Selvam (1993) has shown that Eq. (1) represents the universal algorithm for deterministic chaos in dynamical systems and is expressed in terms of the universal *Feigenbaum's* (1980) *constants a* and *d* as follows. The successive length step growths generating the eddy continuum $OR_OR_1R_2R_3R_4R_5$ analogous to the period doubling route to chaos (growth) is initiated and sustained by the turbulent (fine scale) eddy acceleration $w_*$, which then propagates by the inherent property of inertia of the medium of propagation. Therefore, the statistical parameters *mean*, *variance*, *skewness* and *kurtosis* of the perturbation field in the medium of propagation are given by $w_*, w_*^2, w_*^3$ and $w_*^4$ respectively. The associated dynamics of the perturbation field can be described by the following parameters. The perturbation speed $w_*$ (motion) per second (unit time) sustained by its inertia represents the mass, $w_*^2$ the acceleration or force, $w_*^3$ the angular momentum or potential energy, and $w_*^4$ the spin angular momentum, since an eddy motion has an inherent curvature to its trajectory.

It is shown that *Feigenbaum's* constant *a* is equal to (Selvam, 1993)

$$a = \frac{W_2 R_2}{W_1 R_1} \tag{8}$$

In Eq. (8) the subscripts 1 and 2 refer to two successive stages of eddy growth. *Feigenbaum's* constant *a* as defined above represents the steady state emergence of fractional *Euclidean* structures. Considering dynamical eddy growth processes, *Feigenbaum's* constant *a* also represents the steady state fractional outward mass dispersion rate and $a^2$ represents the energy flux into the environment generated by the persistent primary perturbation $W_1$. Considering both clockwise and counterclockwise rotations, the total energy flux into the environment is equal to $2a^2$. In statistical terminology, $2a^2$ represents the variance of fractal structures for both clockwise and counterclockwise rotation directions.

The probability of occurrence $P_{tot}$ of fractal domain $W_1R_1$ in the total larger eddy domain $W_nR_n$ in any (irrespective of positive or negative) direction is equal to

$$P_{tot} = \frac{W_1 R_1}{W_n R_n} = \tau^{-2n}$$

Therefore the probability *P* of occurrence of fractal domain $W_1R_1$ in the total larger eddy domain $W_nR_n$ in any one direction (either positive or negative) is equal to

$$P = \left(\frac{W_1 R_1}{W_n R_n}\right)^2 = \tau^{-4n} \tag{9}$$

The *Feigenbaum's* constant *d* is shown to be equal to (Selvam, 1993)

$$d = \frac{W_2^4 R_2^3}{W_1^4 R_1^3} \tag{10}$$

Eq. (10) represents the fractional volume intermittency of occurrence of fractal structures for each length step growth. *Feigenbaum's* constant $d$ also represents the relative spin angular momentum of the growing large eddy structures as explained earlier.

Eq. (1) may now be written as

$$2\frac{W^2 R^2}{w_*^2 (dR)^2} = \pi \frac{W^4 R^3}{w_*^4 (dR)^3} \tag{11}$$

In Eq. (11) $dR$ equal to $r$ represents the incremental growth in radius for each length step growth, i.e., $r$ relates to the earlier stage of eddy growth.

The Feigenbaum's constant $d$ represented by $R/r$ is equal to

$$d = \frac{W^4 R^3}{w_*^4 r^3} \tag{12}$$

For two successive stages of eddy growth

$$d = \frac{W_2^4 R_2^3}{W_1^4 R_1^3} \tag{13}$$

From Eq. (1)

$$W_1^2 = \frac{2}{\pi} \frac{r}{R_1} w_*^2$$
$$W_2^2 = \frac{2}{\pi} \frac{r}{R_2} w_*^2 \tag{14}$$

Therefore

$$\frac{W_2^2}{W_1^2} = \frac{R_1}{R_2} \tag{15}$$

Substituting in Eq. (13)

$$d = \frac{W_2^4 R_2^3}{W_1^4 R_1^3} = \frac{W_2^2}{W_1^2} \frac{W_2^2 R_2^3}{W_1^2 R_1^3} = \frac{R_1}{R_2} \frac{W_2^2 R_2^3}{W_1^2 R_1^3} = \frac{W_2^2 R_2^2}{W_1^2 R_1^2} \tag{16}$$

The Feigenbaum's constant $d$ represents the scale ratio $R_2/R_1$ and the inverse of the Feigenbaum's constant $d$ equal to $R_1/R_2$ represents the probability $(Prob)_1$ of occurrence of length scale $R_1$ in the total fluctuation length domain $R_2$ for the first eddy growth step as given in the following

$$(Prob)_1 = \frac{R_1}{R_2} = \frac{1}{d} = \frac{W_1^2 R_1^2}{W_2^2 R_2^2} = \tau^{-4} \tag{17}$$

In general for the $n^{th}$ eddy growth step, the probability $(Prob)_n$ of occurrence of length scale $R_1$ in the total fluctuation length domain $R_n$ is given as

$$(Prob)_n = \frac{R_1}{R_n} = \frac{W_1^2 R_1^2}{W_n^2 R_n^2} = \tau^{-4n} \tag{18}$$

The above equation for probability $(Prob)_n$ also represents, for the $n^{th}$ eddy growth step, the following statistical and dynamical quantities of the growing large eddy with respect to the initial perturbation domain: (i) the statistical relative variance of fractal structures, (ii) probability of occurrence of fractal domain in either positive or negative direction, and (iii) the dynamical relative energy flux.

Substituting the *Feigenbaum's constants a* and *d* defined above (Eqs. 8 and 10), Eq. (11) can be written as

$$2a^2 = \pi d \tag{19}$$

In Eq. (19) $\pi d$, the relative volume intermittency of occurrence contributes to the total variance $2a^2$ of fractal structures.

In terms of eddy dynamics, the above equation states that during each length step growth, the energy flux into the environment equal to $2a^2$ contributes to generate relative spin angular momentum equal to $\pi d$ of the growing fractal structures.

It was shown at Eq. (5) above that the steady state emergence of fractal structures in fluid flows is equal to $1/k$ ($=\tau^2$) and therefore the *Feigenbaum's constant a* is equal to

$$a = \tau^2 = \frac{1}{k} = 2.62 \tag{20}$$

(d) The power spectra of fluctuations in fluid flows can now be quantified in terms of universal *Feigenbaum's constant a* as follows.

The normalized variance and therefore the statistical probability distribution is represented by (from Eq. 9)

$$P = a^{-2t} \tag{21}$$

In Eq. (21) $P$ is the probability density corresponding to normalized standard deviation $t$. The graph of $P$ versus $t$ will represent the power spectrum. The slope $S$ of the power spectrum is equal to

$$S = \frac{dP}{dt} \approx -P \tag{22}$$

The power spectrum therefore follows inverse power law form, the slope decreasing with increase in $t$. Increase in $t$ corresponds to large eddies (low frequencies) and is consistent with observed decrease in slope at low frequencies in dynamical systems.

The steady state emergence of fractal structures for each length step growth for any one direction of rotation (either clockwise or anticlockwise) is equal to

$$\frac{a}{2} = \frac{\tau^2}{2}$$

since the corresponding value for both direction is equal to *a* (Eqs. 5 and 20 ).

The emerging fractal space-time structures have moment coefficient of kurtosis given by the fourth moment equal to

$$\left(\frac{\tau^2}{2}\right)^4 = \frac{\tau^8}{16} = 2.9356 \approx 3$$

The moment coefficient of skewness for the fractal space-time structures is equal to zero for the symmetric eddy circulations. Moment coefficient of kurtosis equal to 3 and moment coefficient of skewness equal to *zero* characterize the statistical normal distribution. The model predicted power law distribution for fractal fluctuations is close to the Gaussian distribution.

(e) The relationship between *Feigenbaum's constant a* and power spectra is derived in the following.

The steady state emergence of fractal structures is equal to the *Feigenbaum's constant a* (Eqs. 5 and 20). The relative variance of fractal structure which also represents the probability *P* of occurrence of bidirectional fractal domain for each length step growth is then equal to $1/a^2$. The normalized variance $\frac{1}{a^{2n}}$ will now represent the statistical probability density for the $n^{th}$ step growth according to model predicted quantumlike mechanics for fluid flows. Model predicted probability density values *P* are computed as

$$P = \frac{1}{a^{2n}} = \tau^{-4n} \qquad (23)$$

or

$$P = \tau^{-4t} \qquad (24)$$

In Eq. (24) *t* is the normalized standard deviation (Eq. 6). The model predicted *P* values corresponding to normalised deviation *t* values less than 2 are slightly less than the corresponding statistical normal distribution values while the *P* values are noticeably larger for normalised deviation *t* values greater than 2 (Table 1 and Fig. 4) and may explain the reported *fat tail* for probability distributions of various physical parameters (Buchanan, 2004). The model predicted *P* values plotted on a linear scale (Y-axis) shows close agreement with the corresponding statistical normal probability values as seen in Fig.4 (left side). The model predicted *P* values plotted on a logarithmic scale (Y-axis) shows *fat tail* distribution for normalised deviation *t* values greater than 2 as seen in Fig.4 (right side).

Table 1: Model predicted and statistical normal probability density distributions

| growth step | normalized deviation | cumulative probability densities (%) | |
|---|---|---|---|
| $n$ | $t$ | model predicted $P = \tau^{-4t}$ | statistical normal distribution |
| 1 | 1 | 14.5898 | 15.8655 |
| 2 | 2 | 2.1286 | 2.2750 |
| 3 | 3 | 0.3106 | 0.1350 |
| 4 | 4 | 0.0453 | 0.0032 |
| 5 | 5 | 0.0066 | ≈ 0.0 |

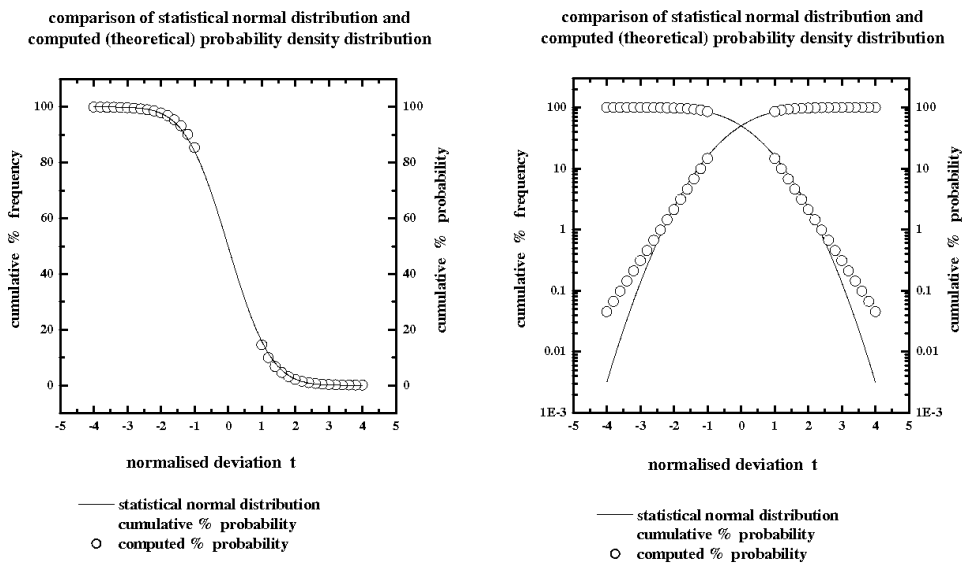

Figure 4: Comparison of statistical normal distribution and computed (theoretical) probability density distribution. The same figure is plotted on the right side with logarithmic scale for the probability axis (Y-axis) to show clearly that for normalized deviation $t$ values greater than 2 the computed probability densities are greater than the corresponding statistical normal distribution values.

## 4. Data and analysis

### 4.1 Data sets used for the study

i. Seven data sets of daily percentage changes in Dow Jones Index for successive 10-year periods starting from the year 1930 to the year 2000. Dow Jones Index values were obtained from Dow Jones Industrial Average History File. Data from: Department of Statistics at Carnegie Mellon Univ., (http://www.stat.cmu.edu/cmu-stats) at the data link http://www.analyzeindices.com/dowhistory/djia-100.txt. The number of trading days in the successive 10-year periods from 1930 to 2000 is shown in Fig. 5a.

ii. The Human chromosomes 1 to 22, x, y DNA base sequence was obtained from the entrez Databases, Homo sapiens Genome (build 36 Version 1) at http://www.ncbi.nlm.nih.gov/entrez. Contiguous data sets, each containing a minimum of 70 000 base pairs were chosen for the study for the chromosomes 1 to 22, x, y. The total number of contiguous data sets, each containing a minimum of 70 000 base pairs, chosen for the study are given in Fig. 6a for the chromosomes 1 to 22, x ,y. The number of times base C and also base G, i.e. (C+G), occur in successive blocks of 10 bases were determined in successive length sections of 70000 base pairs giving a C+G frequency distribution series of 7000 values for each data set.

iii. The draft sequence of Takifugu rubripes (Puffer fish) genome assembly release 4 was obtained from "The Fugu Informatics Network" (ftp://fugu.biology.qmul.ac.uk/pub/fugu/scaffolds_4.zip) at School of Biological & Chemical Sciences, Queen Mary, University of London. The individual contigs sizes range from 2-1100Kbp, nearly half the genome in just 100 scaffolds, 80% of the genome in 300 scaffolds. Non-overlapping DNA sequence lengths without breaks (N values) were chosen and then grouped in two categories A and B; category A consists of DNA lengths greater than 3Kbp (kilo base pairs) but less than 30Kbp and category B consists of DNA sequence lengths greater than 30Kbp. For convenience in averaging the results, the Category A data lengths were grouped into 7 groups and category B data into 18 groups such that the total number of bases in each group is equal to about 12Mbp. The average, standard deviation, maximum, minimum and median DNA lengths (bp) for each group in the two data categories A and B are shown in Fig. 7a. It is seen that the mean is close to the median and almost constant for the different data groups, particularly for category B (DNA length > 30kbp). The number of times base C and also base G, i.e., (C+G), occur in successive blocks of 10 bases were determined in the DNA length sections giving a C+G frequency distribution series of 300 to 3000 values in category A and more than 3000 for category B.

**4.2 Analyses and results**

Each data set was represented as the frequency of occurrence $f(i)$ in a suitable number $n$ of class intervals $x(i)$, $i=1, n$ covering the range of values from *minimum* to the *maximum* in the data set. The class interval $x(i)$ represents dataset values in the range $x(i) \pm \Delta x$, where $\Delta x$ is a constant. The average $av$ and standard deviation $sd$ for the data set is computed as

$$av = \frac{\sum_{1}^{n}[x(i) \times f(i)]}{\sum_{1}^{n} f(i)}$$

$$sd = \frac{\sum_{1}^{n}\{[x(i) - av]^2 \times f(i)\}}{\sum_{1}^{n} f(i)}$$

The average and standard deviation values are given in Figs. 5a, 6a and 7a respectively for Dow Jones Index daily percentage changes, Human chromosomes 1 to 22, x, y DNA base CG/10bp distribution and Takifugu rubripes (Puffer fish) DNA base CG/10bp distribution.

The *normalized deviation t* values for class intervals $t(i)$ were then computed as

$$t(i) = \frac{x(i) - av}{sd}$$

The cumulative percentage probabilities of occurrence *cmax*(*i*) and *cmin*(*i*) were then computed starting respectively from the maximum (*i=n*) and minimum (*i=1*) class interval values as follows.

$$cmax(i) = \frac{\sum_{n}^{i}[x(i) \times f(i)]}{\sum_{1}^{n}[x(i) \times f(i)]} \times 100.0$$

$$cmin(i) = \frac{\sum_{1}^{i}[x(i) \times f(i)]}{\sum_{1}^{n}[x(i) \times f(i)]} \times 100.0$$

The cumulative percentage probability values *cmax*(*i*) and *cmin*(*i*) plotted with respect to corresponding *normalized deviation t(i)* values with both linear and logarithmic scale for the probability axis are shown in (i) Fig. 5b for Dow Jones Index daily percentage changes, (ii) Fig. 6.1b for Human chromosomes 1 to 6 DNA base CG/10bp distributions, (iii) Fig. 6.2b for Human chromosomes 7 to 12 DNA base CG/10bp distributions, (iv) Fig. 6.3b for Human chromosomes 13 to x, y DNA base CG/10bp distributions, (v) Fig. 7b for Takifugu rubripes (Puffer fish) DNA base CG/10bp distributions. The figures also contain the statistical normal distribution for comparison. The graphs with the probability axis on logarithmic scale includes the computed theoretical probabilities (Eq. 24) to show clearly the appreciable departure of observed probability densities from the statistical normal distribution at normalised deviation *t* values more than 2.

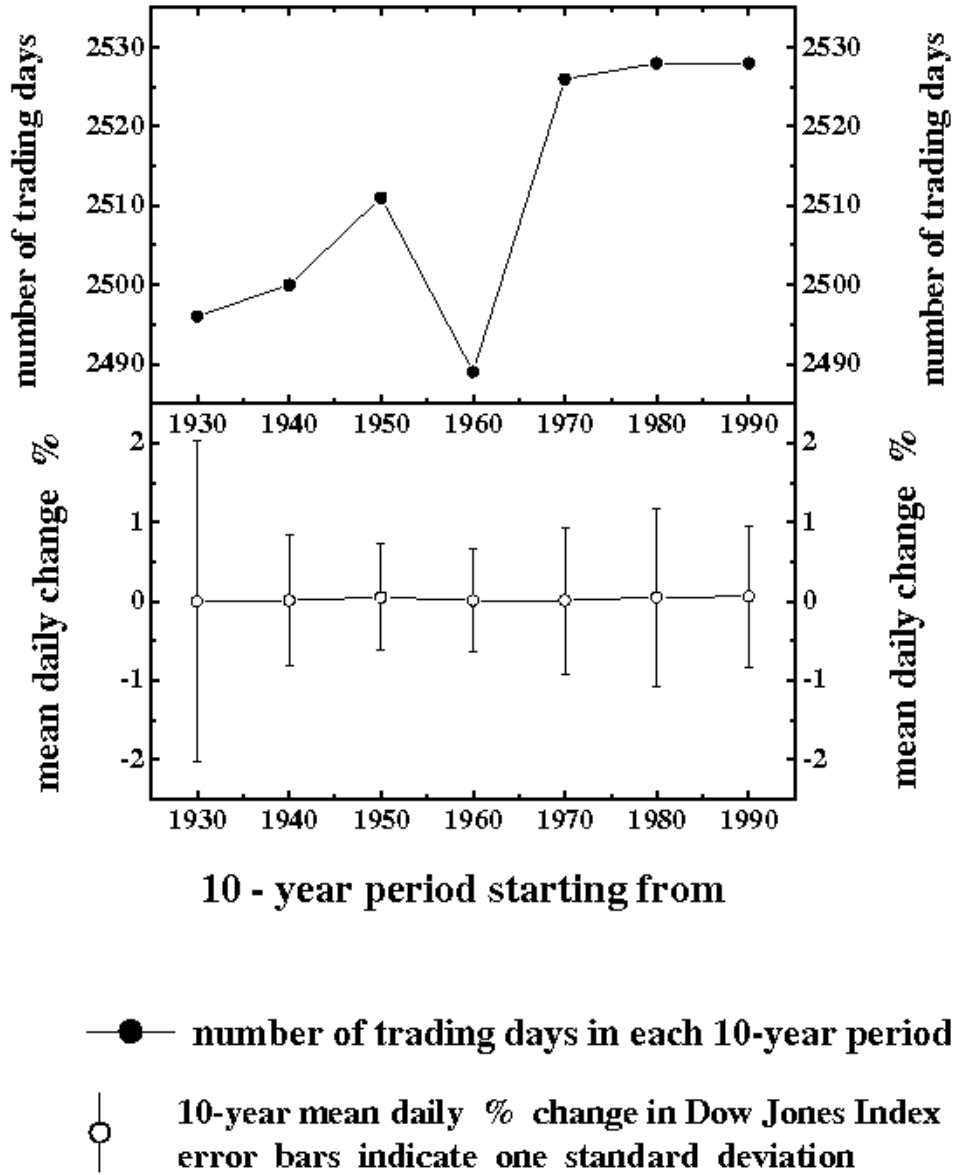

Figure 5a. Details of data sets for Dow Jones Index 10-year mean daily percentage changes for successive 10-year periods from 1930 to 2000.

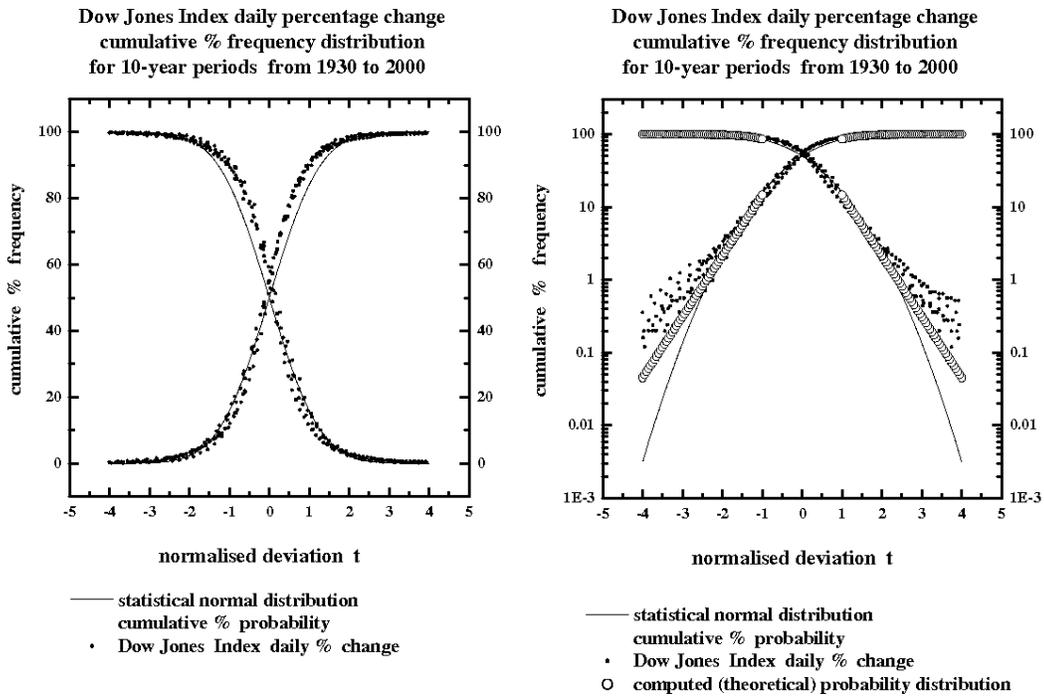

Figure 5b. The cumulative percentage frequency distribution versus the normalized deviation $t$ for Dow Jones Index daily percentage changes for successive 10-year periods from 1930 to 2000, computed starting from both maximum positive and maximum negative $t$ values. The corresponding statistical normal probability density distribution is shown in the figure. The same figure is plotted on the right side with logarithmic scale for probability (Y-axis) and includes computed (theoretical) probability densities.

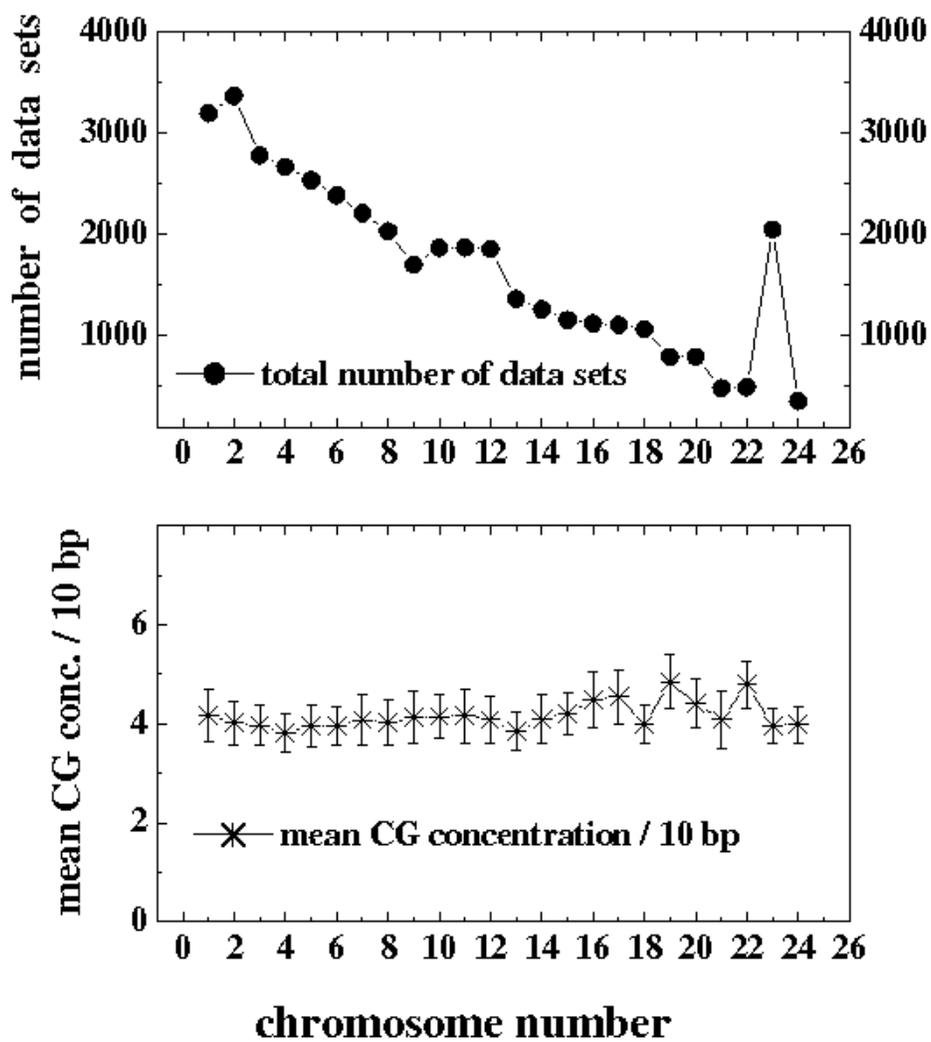

Figure 6a: Details of Human chromosomes 1 to 22, X, Y data sets used in the study

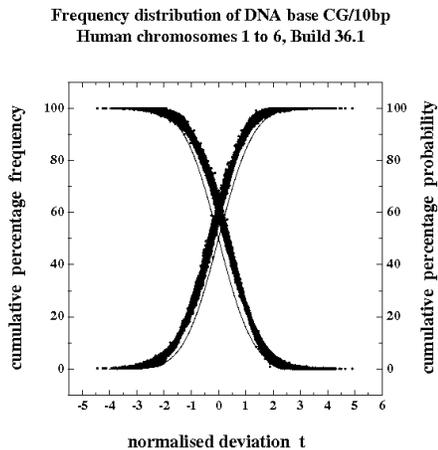 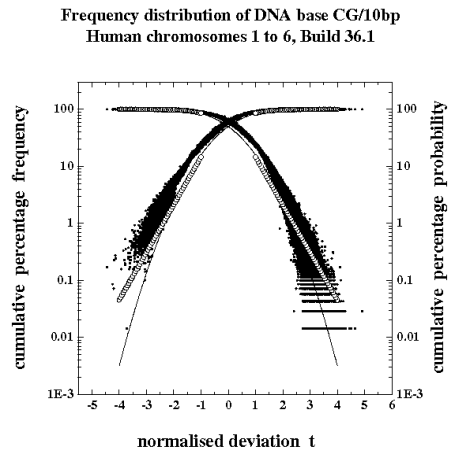

Figure 6.1b: The cumulative percentage frequency distribution versus the normalized deviation $t$ for DNA base CG/10bp in Human chromosomes 1 to 6 (Build 36.1) computed starting from both maximum positive and maximum negative $t$ values. The corresponding statistical normal probability density distribution is shown in the figure. The same figure is plotted on the right side with logarithmic scale for probability (Y-axis) and includes computed (theoretical) probability densities.

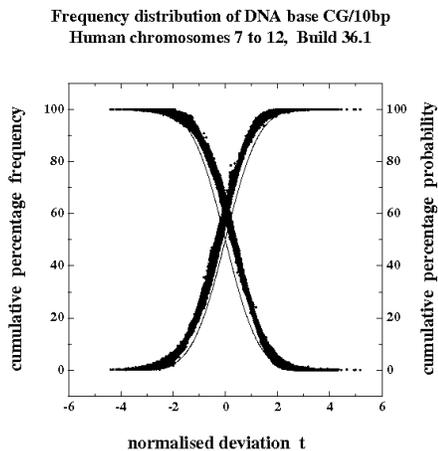 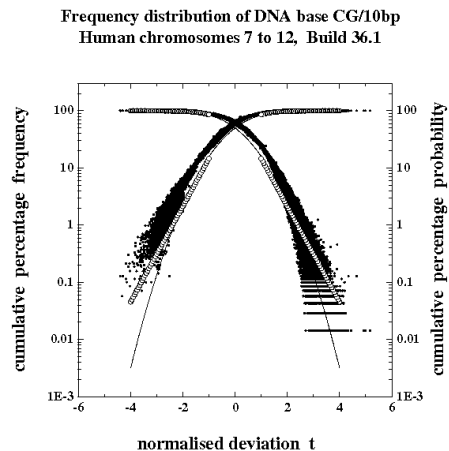

Figure 6.2b: The cumulative percentage frequency distribution versus the normalized deviation $t$ for DNA base CG/10bp in Human chromosomes 7 to 12 (Build 36.1) computed starting from both maximum positive and maximum negative $t$ values. The corresponding statistical normal probability density distribution is shown in the figure. The same figure is plotted on the right side with logarithmic scale for probability (Y-axis) and includes computed (theoretical) probability densities.

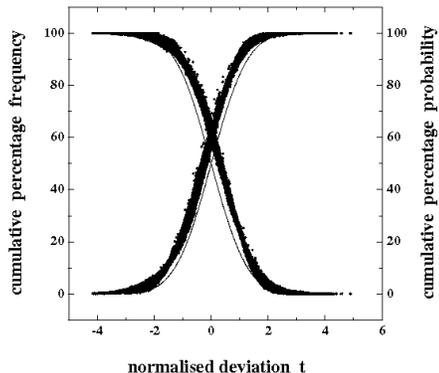 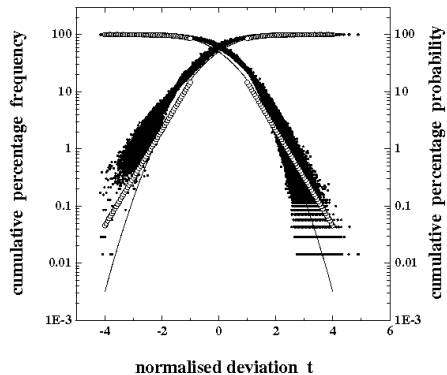

Figure 6.3b: The cumulative percentage frequency distribution versus the normalized deviation $t$ for DNA base CG/10bp in Human chromosomes 13 to 22, X, Y (Build 36.1) computed starting from both maximum positive and maximum negative $t$ values. The corresponding statistical normal probability density distribution is shown in the figure. The same figure is plotted on the right side with logarithmic scale for probability (Y-axis) and includes computed (theoretical) probability densities.

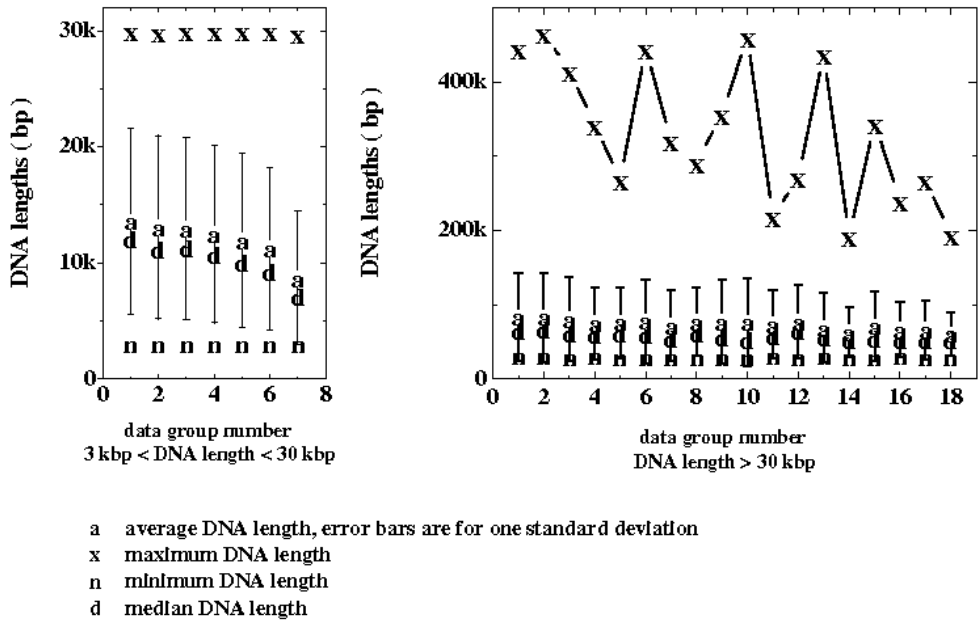

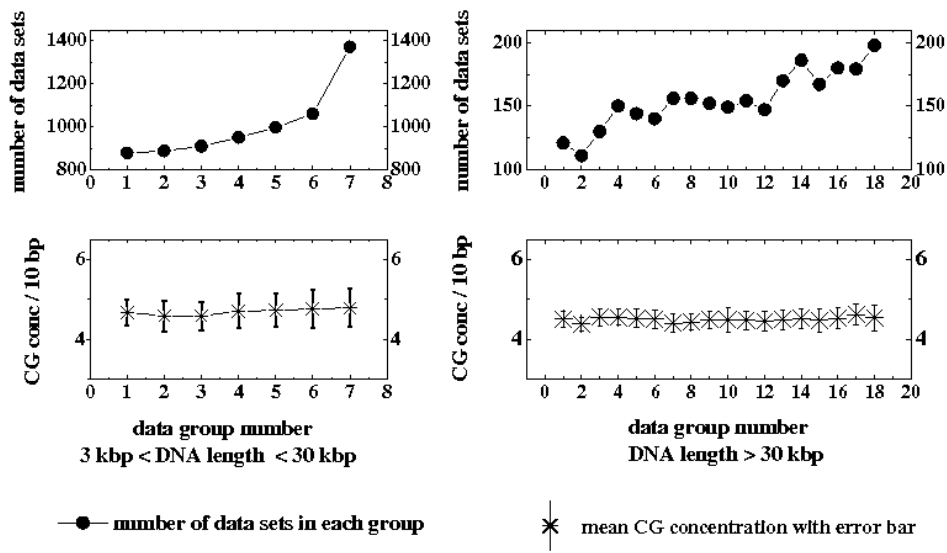

Fig. 7a. Takifugu rubripes (Puffer fish) DNA base CG/10bp concentration. Details of data sets and averaged results .

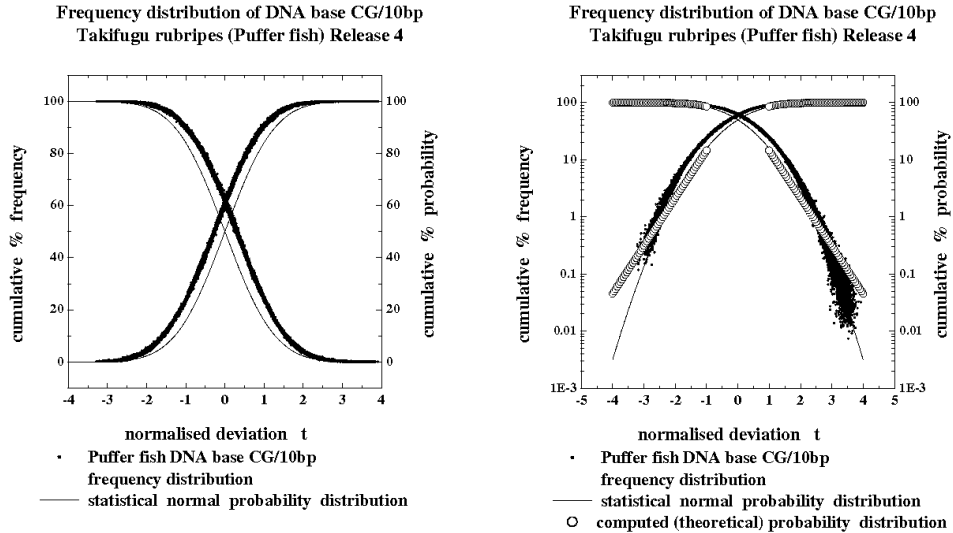

Figure 7b: The cumulative percentage frequency distribution versus the normalized deviation *t* for DNA base CG/10bp in Takifugu rubripes (Puffer fish) computed starting from both maximum positive and negative *t* values. The corresponding statistical normal probability density distribution is also shown in the figure. The same figure is plotted on the right side with logarithmic scale for probability (Y-axis) and includes computed (theoretical) probability densities.

## 5. Discussion and conclusions

Dynamical systems in nature exhibit selfsimilar fractal fluctuations for all space-time scales and the corresponding power spectra follow inverse power law form signifying long-range space-time correlations identified as self-organized criticality. The physics of self-organized criticality is not yet identified. The Gaussian probability distribution used widely for analysis and description of large data sets is found to significantly underestimate the probabilities of occurrence of extreme events such as stock market crashes, earthquakes, heavy rainfall, etc. Further, the assumptions underlying the normal distribution such as fixed mean and standard deviation, independence of data, are not valid for real world fractal data sets exhibiting a scale-free power law distribution with fat tails. It is important to identify and quantify the fractal distribution characteristics of dynamical systems for predictability studies.

A recently developed general systems theory for fractal space-time fluctuations (Selvam, 1990, 2005, 2007; Selvam and Fadnavis, 1998) shows that the larger scale fluctuation can be visualized to emerge from the space-time averaging of enclosed small scale fluctuations, thereby generating a hierarchy of selfsimilar fluctuations manifested as the observed eddy continuum in power spectral analyses of fractal fluctuations. Such a concept results in inverse power law form $\tau^{-4t}$ incorporating $\tau$, the golden mean and *t*, the normalized deviation, for the space-time fluctuation pattern and also for the power spectra of the fluctuations (Sec. 3). Since the power spectrum (square of eddy amplitude) also represents the probability densities as in the case of quantum systems such as the electron or photon, fractal fluctuations exhibit quantumlike chaos. The predicted distribution is close to the Gaussian distribution for small-scale fluctuations, but exhibits *fat long tail* for large-scale fluctuations. Analysis of extensive data sets of (i) Daily percentage change of Dow Jones Index (ii) Human DNA base CG concentration/10bp (base pairs) (iii) Takifugu rubripes (Puffer fish) DNA base CG concentration/10bp give the following results: (a) the data follow closely, but not exactly the statistical normal distribution in the region of normalized deviations *t* less than 2, the *t* values being computed as equal to (*x-av*)/*sd* where *av* and *sd* denote respectively the mean and standard deviation of the variable *x* (b) For normalized deviations *t* greater than 2, the data exhibit significantly larger probabilities as compared to the normal distribution and closer to the model predicted probability distribution. The general systems theory, originally developed for turbulent fluid flows, provides universal quantification of physics underlying fractal fluctuations and is applicable to all dynamical systems in nature independent of its physical, chemical, electrical, or any other intrinsic characteristic.

## Acknowledgement

The author is grateful to Dr. A. S. R. Murty for encouragement during the course of this study.

## References

Andriani, P. and McKelvey. B., 2007: Beyond Gaussian averages: Redirecting management research toward extreme events and power laws. *Journal of International Business Studies* **38**, 1212–1230.

Auerbach, F., 1913: Das Gesetz Der Bevolkerungskoncentration, *Petermanns Geographische Mitteilungen* **59**, 74–76.

Bak, P. C., Tang, C. and Wiesenfeld, K., 1988: Self-organized criticality. *Phys. Rev. A*. **38**, 364 - 374.

Bouchaud, J. P., Sornette, D., Walter, C. and Aguilar, J. P., 1998: Taming large events: Optimal portfolio theory for strongly fluctuating assets. *International Journal of Theoretical and Applied Finance* **1**(1), 25–41.

Bradley, J. V., 1968: *Distribution-free Statistical Tests*. Englewood Cliffs, Prentice-Hall, N.J.

Buchanan, M., 2004: Power laws and the new science of complexity management. *Strategy and Business Issue* **34**, 70-79.

Clauset, A., Shalizi, C. R. and Newman, M. E. J., 2007: Power-law distributions in empirical data. arXiv:0706.1062v1 [physics.data-an].

Cronbach, L., 1970: *Essentials of Psychological Testing*. Harper & Row, New York.

Estoup, J. B., 1916: *Gammes Stenographiques*, Institut Stenographique de France, Paris.

Fama, E. F., 1965:The behavior of stock-market prices. *Journal of Business* **38**, 34–105.

Feigenbaum, M. J., 1980: Universal behavior in nonlinear systems. *Los Alamos Sci*. **1**, 4-27.

Goertzel, T. and Fashing, J., 1981: The myth of the normal curve: A theoretical critique and examination of its role in teaching and research. *Humanity and Society* **5**, 14-31; reprinted in *Readings in Humanist Sociology*, General Hall, 1986. http://crab.rutgers.edu/~goertzel/normalcurve.htm 4/29/2007

Greene, W. H., 2002: *Econometric Analysis* (5th ed.). Prentice-Hall, Englewood Cliffs, NJ.

Grossing, G., 1989: Quantum systems as order out of chaos phenomena. *Il Nuovo Cimento* **103**B, 497-510.

Gutenberg, B. and Richter, R. F., 1944: Frequency of earthquakes in California. *Bulletin of the Seismological Society of America* **34**, 185–188.

Liebovitch, L. S. and Scheurle, D., 2000: Two lessons from fractals and chaos. *Complexity* **5(4)**, 34–43.

Maddox, J., 1988: Licence to slang Copenhagen? *Nature* **332**, 581.

Maddox, J., 1993: Can quantum theory be understood? *Nature* **361**, 493.

Mandelbrot, B. B., 1963: The variation of certain speculative prices. *Journal of Business* **36**, 394–419.

Mandelbrot, B. B., 1975: *Les Objets Fractals: Forme, Hasard et Dimension*. Flammarion, Paris.

Mandelbrot, B. B. and Hudson, R. L., 2004: *The (Mis)Behavior of Markets: A Fractal View of Risk, Ruin and Reward*. Profile, London.

Montroll, E. and Shlesinger, M., 1984: On the wonderful world of random walks, in J. L. Lebowitz and E. W. Montroll (eds.) *Nonequilibrium Phenomena II, from Stochastic to Hydrodynamics*. North Holland, Amsterdam, pp. 1–121.

Omori, F., 1895: On the aftershocks of earthquakes. *J. Coll. Sci*. **7**, 111.

Pareto, V., 1897: *Cours d'Economie Politique*. Rouge, Paris.

Pearson, K., 1900: *The Grammar of Science* (2nd ed.). Adam and Charles Black, London:

Phillips, T., 2005: *The Mathematical Uncertainty Principle*. Monthly Essays on Mathematical Topics, November 2005, American Mayhematical Society. http://www.ams.org/featurecolumn/archive/uncertainty.html

Rae, A., 1988: *Quantum-Physics: Illusion or Reality?* Cambridge University Press, New York.

Richardson, L. F., 1960: The problem of contiguity: An appendix to statistics of deadly quarrels, In: Von Bertalanffy, L., Rapoport, A., (eds.) *General Systems - Year Book of The Society for General Systems Research* V, pp 139-187, Ann Arbor, MI.

Riley, K. F., Hobson, M. P. and Bence, S. J., 2006: *Mathematical Methods for Physics and Engineering* (3rd ed.). Cambridge University Press, USA.

Ruhla, C. 1992: *The Physics of Chance*. Oxford University Press, pp.217.

Schroeder, M., 1990: *Fractals, Chaos and Power-laws*. W. H. Freeman and Co., N. Y.

Selvam, A. M., 1990: Deterministic chaos, fractals and quantumlike mechanics in atmospheric flows. *Can. J. Phys*. **68**, 831-841. http://xxx.lanl.gov/html/physics/0010046

Selvam, A. M., 1993: Universal quantification for deterministic chaos in dynamical systems. *Applied Math. Modelling* **17**, 642-649. http://xxx.lanl.gov/html/physics/0008010

Selvam A. M., Fadnavis, S., 1998: Signatures of a universal spectrum for atmospheric inter-annual variability in some disparate climatic regimes. *Meteorology and Atmospheric Physics* **66**, 87-112. http://xxx.lanl.gov/abs/chao-dyn/9805028

Selvam, A. M., and Fadnavis, S., 1999: Superstrings, cantorian-fractal spacetime and quantum-like chaos in atmospheric flows. *Chaos Solitons and Fractals* **10**, 1321-1334. http://xxx.lanl.gov/abs/chao-dyn/9806002

Selvam, A. M., Sen, D. and Mody, S. M. S., 2000: Critical fluctuations in daily incidence of acute myocardial infarction. *Chaos, Solitons and Fractals* **11**, 1175-1182. http://xxx.lanl.gov/abs/chao-dyn/9810017

Selvam, A. M. 2001a: Quantum-like chaos in prime number distribution and in turbulent fluid flows. *Apeiron* **8**, 29-64. http://redshift.vif.com/JournalFiles/V08NO3PDF/V08N3SEL.PDF http://xxx.lanl.gov/html/physics/0005067

Selvam, A. M. 2001b: Signatures of quantum-like chaos in spacing intervals of non-trivial Riemann zeta zeros and in turbulent fluid flows. *Apeiron* **8**, 10-40. http://redshift.vif.com/JournalFiles/V08NO4PDF/V08N4SEL.PDF http://xxx.lanl.gov/html/physics/0102028

Selvam, A. M., 2002a: Cantorian fractal space-time fluctuations in turbulent fluid flows and the kinetic theory of gases. *Apeiron* **9**,1-20. http://redshift.vif.com/JournalFiles/V09NO2PDF/V09N2sel.PDF http://xxx.lanl.gov/html/physics/9912035

Selvam, A. M., 2002b: Quantumlike chaos in the frequency distributions of the bases A, C, G, T in Drosophila DNA. *Apeiron* **9**, 103-148. http://redshift.vif.com/JournalFiles/V09NO4PDF/V09N4sel.pdf http://arxiv.org/html/physics/0210068

Selvam, A. M., 2004: Quantumlike Chaos in the Frequency Distributions of the Bases A, C, G, T in human chromosome 1 DNA. *Apeiron* **11**, 134-146. http://redshift.vif.com/JournalFiles/V11NO3PDF/V11N3SEL.PDF http://arxiv.org/html/physics/0211066

Selvam, A. M., 2005: A general systems theory for chaos, quantum mechanics and gravity for dynamical systems of all space-time scales. *ELECTROMAGNETIC PHENOMENA* **5**, No.2(15), 160-176. http://arxiv.org/pdf/physics/0503028

Selvam, A. M., 2007: *Chaotic Climate Dynamics*. Luniver Press, UK.

Steinhardt, P., 1997: Crazy crystals. *New Scientist* **25 Jan**., 32-35.

Townsend, A. A., 1956: *The Structure of Turbulent Shear Flow* (2nd ed.) Cambridge University Press, London, U.K., pp.115-130.

Zipf, G.K., 1949: *Human Behavior and the Principle of Least Effort*. Hafner, New York.